\documentclass[%
 reprint,
 amsmath,amssymb,
 aps,
]{revtex4-2}
\usepackage{subfig}
\usepackage{graphicx}
\usepackage{dcolumn}
\usepackage{bm}

\begin{document}

\preprint{APS/123-QED}

\title{Phase separation dynamics and active turbulence in a binary fluid mixture}

\author{Sohail Ahmed, Zixiang Lin and Zijie Qu}

 \email{E-mail: zijie.qu@sjtu.edu.cn}
\affiliation{%
 Global College, Shanghai Jiao Tong University, Shanghai 200240, People’s Republic of China 
}%

\date{\today}

\begin{abstract}
Active matter, encompassing natural systems, converts surrounding energy to sustain autonomous motion, exhibiting unique non-equilibrium behaviors such as active turbulence and phase separation. In this study, we develop a continuum two-fluid model for a binary mixture of an active nematic and a passive Newtonian fluid, coupling Cahn–Hilliard dynamics for phase separation with Beris–Edwards nematohydrodynamics and two distinct momentum equations connected by viscous drag. A phase field-based lattice Boltzmann method is used to investigate the existence of active turbulence and phase separation in the binary mixture. We find that active stress enhances velocity and vorticity differences between phases, and that increased active concentration promotes stronger inter-fluid coupling. Activity not only amplifies turbulent fluctuations but also arrests domain coarsening, leading to a finite characteristic length scale that decreases with increasing activity.  Key parameters, like active parameter, tumbling parameter and Frank elastic constant, affect the characteristic scale of flow. These results highlight the role of relative motion and drag-mediated momentum transfer in active binary mixtures, providing a framework for studying systems such as bacterial suspensions in polymeric fluids or active emulsions.
\end{abstract}

\maketitle


\section{Introduction}
Research on active fluids has garnered a noticeable attention over the past years with real-world applications \cite{yang-NM-2025, volpe-npjm-2022, GHOSH-NT-2020}. These fluids, which range from biological examples such as bacterial swarms and cellular tissues to synthetic active colloids are complex fluids identified by the presence of an active phase whose individual units self-propel.\cite{Marchetti-RMP-2013,Ramaswamy-ARCMP-2010} Active fluids are prone to self-organization phenomena, thus developing correlated collective movements that can become spatio-temporally chaotic pattern, consisting of fluid jets and swirls referred to as active turbulence.\cite{Wensink-PNAS-2012, Alert-ARCMP-2022} Experimentally, this novel form of turbulence manifests a low Reynolds number phenomenon \cite{Wensink-PNAS-2012, Dombrowski-PRL-2004}. Contrary to the classical turbulence of other fluids, which occurs when fluid is driven by an external force, active fluids are driven by microscopic constituents that transform active energy sources into kinetic energy \cite{Dunkel-PRL-2013, Peter-PNAS-2023}. Continuum hydrodynamic models have been used to study the emergence of turbulence-like patterns.\cite{Alert-ARCMP-2022} The kinetic energy spectrum of such turbulence in some theoretical models displays  universal power-law behaviors depending on whether the wavenumber is greater or less than a typical vortex size. The $k^{-4}$ scaling has been reported in early numerical studies \cite{mukherjee-NP-2023, Giomi-PRX-2015}, while subsequent high-resolution Stokes flow simulations observed $k^{-1}$ behavior in certain parameter regimes \cite{alert-NP-2020}. These scaling laws remain an active area of investigation.

Active nematic fluids \cite{Doostmohammadi-NC-2018} are a particularly interesting type that contain highly elongated polar interacting units. Systems composed of vibrated monolayers of cylindrical rods,\cite{vijay-Science-2007} actin filaments,\cite{Nitin-SA-2018} and certain types of bacteria \cite{He-PNAS-2019} suspended in a fluid layer are examples of active nematic fluids. Experiments have largely been motivated by a desire to comprehend the physics of biological systems, and theoretical and computational advancements have been built upon the foundational knowledge of passive liquid crystals.\cite{THAMPI-CPCIS-2022} Among the classical continuum mechanics models for describing nematic liquid crystals, the Beris-Edwards model is the most comprehensive, replacing the director field with a Q-tensor field \cite{Qian-PRE-1998} and thus allowing for a variable degree of order in the material. This model has become a cornerstone for studying many active nematic systems, successfully reproducing key experimental phenomena such as active turbulence.\cite{Giomi-PRX-2015} When extended to binary mixtures, the Beris-Edwards framework provides a powerful tool to investigate how activity-driven flows interact with phase separation in systems composed of an active nematic and a passive isotropic fluid.\cite{Bhattacharyya-PRL-2023} In such systems, the interplay between activity, hydrodynamics, and composition leads to rich non-equilibrium phenomena, including the arrest of phase separation and the formation of active turbulent states central themes explored in this work.

Phase separation, a concept in physical chemistry, refers to dynamic processes where a homogeneous system spontaneously splits into discrete phases with various compositions and physical properties as a result of changes in external conditions and composition characteristics. Clifford {\em et al.}\cite{Clifford-Science-2009} found in 2009 that RNA and protein-based granules in Caenorhabditis elegans embryos form droplets through phase separation. This gave rise to the idea of phase separation in biology, which describes how biomolecules such as proteins, lipids, and nucleic acids interact with one another through multiple bonds to form membrane-less organelles or bimolecular condensates in a uniform environment.\cite{Clifford-PNAS-2011} Motility-induced phase separation (MIPS) is one of the example of active phase separation, in which self-propelled particles can get trapped in highly dense region, forming a dense phase and a dilute phase.\cite{Cates-ARCMP-2015, Mondal-SM-2025} MIPS has been thoroughly investigated in both single-component systems and binary mixtures by theoretical and numerical approaches.\cite{Wittkowski-NC-2014, Jeong-SM-2024} Continuum models of active matter have been used to study the phase separation in active Brownian particles \cite{Stenhammar-PRL-2013}, cellular aggregate \cite{Kuan-PRL-2021}, self-propelled particles.\cite{Tiribocchi-PRL-2015} Some other representative studies where Cahn-Hilliard model used for the phase separation can be found in the works of Yin and Mahadevan \cite{yin-PRL-2023}, Speck \cite{Speck-PRL-2014} and Saha {\em et al.}\cite{Saha-PRX-2020}.

Recent few studies have already demonstrated that two-fluid formulations can capture the coupled dynamics of an active nematic component and a passive isotropic component. Assante {\em et al.} \cite{Assante-SM-2023} combined Cahn–Hilliard dynamics, Beris–Edwards nematohydrodynamics, and hydrodynamic flow to investigate microphase separation and turbulent-like behavior in inhomogeneous active nematics. Extending their work, Bhattacharyya and Yeomans \cite{Bhattacharyya-PRE-2024} developed a continuum two-fluid model incorporating distinct velocities, compositional dynamics, and interphase drag, showing that activity-driven flows can induce microphase separation in mixed active–passive systems. These works establish that coupling a composition field with nematic ordering and hydrodynamics produces rich non-equilibrium structure and flow. However, the detailed mechanisms through which interphase drag, activity strength, and the tumbling parameter collectively govern the relative motion, energy transfer, and spectral characteristics in such two-fluid active systems remain less explored. To address this gap, we investigate a two-fluid active/passive mixture in which Cahn–Hilliard phase separation governs the spatial distribution of both components, Beris–Edwards nematohydrodynamics describes the orientational order and active stresses, and two distinct momentum equations coupled by viscous drag resolve the relative motion between the components. Our aim is to analyze how activity, tumbling versus shear-aligning behavior, and overall concentration influence the flow statistics, spatial correlations, and kinetic-energy distributions of both fluids. By integrating phase separation, two-fluid hydrodynamics, and active nematic stresses, this work provides a focused extension and comprehensive numerical investigation within the established family of two-fluid active nematic models.
\section{Model}
We model a mixture of an active fluid and passive fluid. Each component has density, velocity and viscosity. The Cahn-Hilliards equation is considered to describe the  phase separation dynamics, which is written as:\cite{Dadvand-POF-2021, Yue-JFM-2004}  
\begin{equation}
\frac{\partial \phi}{\partial t}+\nabla \cdot (\phi u_{c}) = M \nabla^{2} (\mu), \quad \mu =\frac{\delta \mathcal{F}}{\delta \phi},  \label{CH-Equation}
\end{equation} 
where  $\phi$ is the concentration of active fluid, ${u_{c}}=\phi u_{1}+(1-\phi)u_{2}$ represents the barycentric (mixture) velocity of the active phase and passive phase, and the non-negative mobility coefficient $M$ is considered often as either a constant or a concentration-dependent variable. $\mathcal{F}$ is the Landau-Ginzburg variation free-energy functional\cite{Yue-JFM-2004} and the chemical potential $\mu$ is defined as the variational derivation of free energy functional with respect to $\phi$ as given in \cite{Yue-JFM-2004}:
\begin{equation}
\mu =\frac{\delta \mathcal{F}}{\delta \phi} = f'(\phi) - \epsilon^{2}\nabla^{2}\phi. 
\end{equation}
The interface thickness parameter $\epsilon$ is a key component of the Cahn-Hilliard equation, ensuring that the interface between different phases is smooth and well-defined. It plays a crucial role in regularizing the interface, controlling its width, and penalizing spatial variations in the concentration field. In the context of two phase flow, $\epsilon$ is essential for accurately modeling the complex dynamics of phase separation and concentration gradients, contributing to the overall understanding of active matter suspension behavior. And $f(\phi)=\phi^{2}(1-\phi)^{2}$ is the double-well potential, enforcing $\phi \in [0,1 ]$. From the above equations we then write:\cite{Dadvand-POF-2021, Yue-JFM-2004}
\begin{equation}
\frac{\partial \phi}{\partial t}+\nabla \cdot (\phi u_{c}) = M \nabla^{2} (f'(\phi) - \epsilon^{2}\nabla^{2}\phi)  \label{CH-Equation-1}.
\end{equation} 
The microscopic momentum equations for each fluid is written as:
\begin{align}
&\phi \rho_{1}\left(\frac{\partial u_{1}}{\partial t}+ u_{1}\cdot \nabla u_{1}\right) = -\phi \nabla P_{1} + \nabla \cdot (\phi \sigma_{1}) 
 \nonumber\\&+F_{drag}+F_{active}+ F_{elastic} + \phi F_{ch}, 
       \label{active phase}
\end{align}
\begin{align}
&\phi_{2}\rho_{2}\left(\frac{\partial u_{2}}{\partial t}+ u_{2}\cdot \nabla u_{2}\right) = -\phi_{2} \nabla P_{2} + \nabla \cdot (\phi_{2} \sigma_{2}) \nonumber\\&-F_{drag}+\phi_{2}F_{ch}, \label{Fluid-Phase}
\end{align}
where $\phi_{2}=(1-\phi)$ is the concentration of passive fluid and $u_{i=1,2}$,  $P_{i=1,2}$,  $\rho_{i=1,2}$, represent 
the velocity, pressure, density of each fluid respectively. For both component we have used a Newtonian viscous stress as defined by Yue {\em et al.}\cite{Yue-JFM-2004}, 
\begin{equation}
    \sigma_{i} = \eta_{i}(\nabla u_{i}+\nabla u_{i}^{T}),\quad i = 1,2.
\end{equation}
Clearly, the viscosity $ \eta_{i=1,2}$  are considered different in this model. While active stress dominates in high-activity regimes, elastic stresses arising from distortions in the nematic ordering can contribute significantly, particularly near topological defects and at low to moderate activities \cite{Doostmohammadi-NC-2018, Thampi-EPL-2015}. Following the complete Beris-Edwards formulation \cite{Saghatchi-IJNME-2023}, we therefore include the full elastic stress tensor in our momentum equations. This ensures that our model captures the interplay between active forcing and elastic restoring forces across all activity regimes. It should be noted that, we have considered four forces namely, the drag force $F_{drag}$, active stress $F_{active}$, elastic stress $F_{elastic}$  and capillary stress $F_{ch}$ related to  active fluid and drag force and capillary stress $F_{ch}$ for the passive fluid. A viscous drag between the components of the fluids $F_{drag}= \gamma \phi (1-\phi) (u_{2}-u_{1})$,\cite{Bhattacharyya-PRE-2024} where $\gamma$ is the momentum transfer coefficient between both fluids. The active stress for an active fluid system is known to take the form of $\zeta$$\textbf{Q}$, with the prominent contribution to active stresses being proportional to the nematic tensor $\textbf {Q}$. $\zeta$ is a phenomenological parameter that presents the activity strength in active fluids being negative for contractile systems, such as (bacteria pull, less common in bacterial suspensions) and positive for extensile systems (bacteria push along their orientation, typical for swimming bacteria like Bacillus subtilis). The factor $\phi$ ensures that this stress is localized to active phase, which form via phase separation (Cahn-Hilliard dynamics). The divergence  $F_{active}=\nabla \cdot (\zeta \phi \textbf{Q})$ translates this stress into a force that accelerates the velocity of the active phase $u_{1}$, interacting with other forces such as drag and interfacial tension. In addition to active forcing, the distortion of the nematic field generates an elastic stress. 
\[\begin{aligned}
F_{\mathrm{elastic}} = & -\lambda \mathbf{H} \cdot \left(\mathbf{Q} + \frac{\mathbf{I}}{2}\right) 
- \lambda \left(\mathbf{Q} + \frac{\mathbf{I}}{2}\right) \cdot \mathbf{H} \\
& + 2\lambda \left(\mathbf{Q} + \frac{\mathbf{I}}{2}\right) (\mathbf{Q}:\mathbf{H}) \\
& - (\nabla \mathbf{Q}) \cdot \frac{\partial \mathcal{F}}{\partial \nabla \mathbf{Q}} 
+ \mathbf{Q} \cdot \mathbf{H} - \mathbf{H} \cdot \mathbf{Q}. \quad
\end{aligned}\] \label{elastic stress}
This elastic stress captures the passive response of the nematic to flow and distortions, and is retained alongside the active stress ${F}_{active}$ to ensure a complete description of the nematohydrodynamics. $\lambda$ is the tumbling parameter, which governs how nematic directors respond to shear. For $|\lambda|<1$, the director tends to tumble under shear, continually rotating rather than aligning with the flow. For $|\lambda|>1$, the system exhibits flow-aligning behavior, where the director reaches a steady orientation relative to the flow. The value $|\lambda|=1$ corresponds to the onset of tumbling-to-aligning transition. Mathematically, $\lambda$ determines the character of objective time derivative of $\textbf{Q}$. $\Gamma$ is the rotational diffusion coefficient and $\Gamma \mathbf{H}$ is relaxational dynamics of the nematic tensor to the minimum of free energy and defined as 
\begin{equation}
    \textbf{H} = -\frac{\delta \textbf{F}}{\delta \textbf{Q}} +(\textbf{I}/2)Tr \frac{\delta \textbf{F}}{\delta \textbf{Q}},  \label{molecular field}
\end{equation}
where Tr denotes the tensorial trace. The Helmholtz free energy\cite{Doostmohammadi-NC-2018} is typically taken as:

\begin{equation}
\textbf{F} = \frac{A}{2} \mathrm{Tr}(Q^2) + \frac{B}{3} \mathrm{Tr}(Q^3) + \frac{C}{4} \mathrm{Tr}(Q^4) + \frac{K}{2} \|\nabla Q\|^2, \label{Total energy}
\end{equation}
and the coefficients A, B, and C are material parameters, and the final term K is elastic constant. And the nematic order parameter $\textbf{Q}$ is presented by a nematic tensor field as: 
 \begin{equation}
     \textbf{Q} =2 S\left(\mathbf{p}\mathbf{p}- \frac{1}{2} \textbf{I}\right).
 \end{equation}
Where, $0\leq S \leq 1$ is the magnitude, $\textbf{pp}$ is the alignment axis of the
nematic ordering, and $\textbf{I}$ is the identity matrix, ensuring that $\textbf{Q}$ is traceless and symmetric, characteristic of nematic liquid crystals. Apart from the active stress, due to an interface between two phases gives rise to another stresses in the flow field. To capture additional stresses, the back-coupling from the concentration of active phase $\phi$ to the fluid equation is inserted through capillary stresses.\cite{Cates-JFM-2018} It can be demonstrated by computing the capillary stress divergence that the associated force field reduces to \cite{Romain-arXiv-2021, Cates-JFM-2018, Yue-JFM-2004}
\begin{equation}
F_{ch}=\sigma_{CH} = \mu \nabla \phi-\nabla (\phi \mu). 
\end{equation} 
The dynamics nematic tensor  $\textbf{Q}$ can be represented by the equation of nematodynamic\cite{Doostmohammadi-NC-2018}
\begin{align}
    \frac{\partial \mathbf{Q}}{\partial t} + \mathbf{u}_1 \cdot \nabla \mathbf{Q} = \mathbf{S}(\nabla \mathbf{u}_1, \mathbf{Q})+\Gamma \mathbf{H}. \label{eq:beris_edwards}
\end{align}
The equation is well known as Beris-Edwards equation, S accounts for the response of the orientational order to the extensional and rotational components of the velocity gradient and is described for the nematic in \cite{Doostmohammadi-NC-2018} as:
\begin{align}
&\mathbf{S}(\nabla \mathbf{u}_1, \mathbf{Q}) =(\lambda \textbf{E} + \Omega)\cdot \left(\textbf{Q}+\frac{\textbf{I}}{2}\right)+\left(\textbf{Q}+\frac{\textbf{I}}{2}\right)\cdot\left(\lambda \textbf{E}-\Omega\right)\\ & \nonumber-2 \lambda \left(\textbf{Q}+\frac{\textbf{I}}{2}\right)\left(\textbf{Q}:\nabla u_{1}\right).
\end{align}
Where $\Omega$ is the vorticity tensor and $\textbf{E}$ is the rate of strain tensor in the fluid\cite{Bhattacharyya-PRL-2023}. 
\section{Numerical Simulation}
\begin{figure}[t]
\centering
\includegraphics[width=0.5\textwidth]{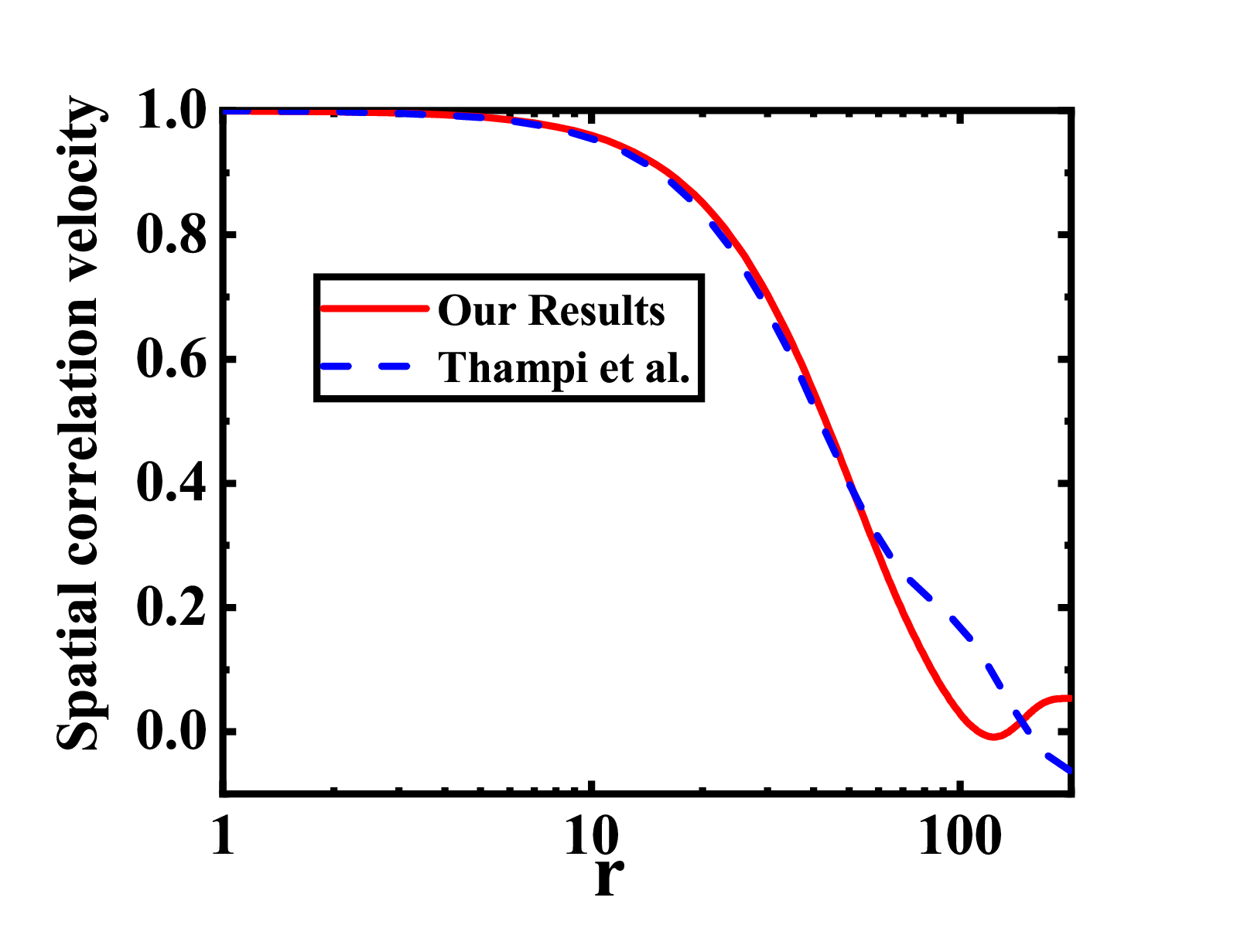}
\caption{Validation of our results with Thampi {\em et al.} \cite{Thampi-PTSAMPES-2014}, for spatial correlation velocity.}\label{Fig01}
\end{figure}

We solve the governing equations described in Section 2 using a hybrid numerical framework combining the lattice Boltzmann method (LBM) for hydrodynamics with finite-difference schemes for the phase-field and nematic order parameter dynamics. The fluid equations are solved on a two-dimensional D2Q9 lattice using a single-relaxation-time Bhatnagar--Gross--Krook (BGK) collision operator \cite{GUO-JCP-2000}, with forcing terms incorporated through the standard Guo forcing scheme \cite{Fei-PRE-2017} to ensure second-order accuracy.  Simulations are performed on a square computational domain of size $256 \times 256$ with periodic boundary conditions applied in both spatial directions for all fields. The spatial and temporal discretizations are chosen as $\Delta x = 1.0$ and $\Delta t = 0.02$, respectively, and the system is evolved for $5 \times 10^4$ time steps to capture the phase separation dynamics and subsequent evolution. The initial conditions consist of a homogeneous phase field with small random perturbations to trigger phase separation, while the velocity fields are initialized to zero, such that $\mathbf{u}_1 = \mathbf{u}_2 = 0$, $\phi(x,y,0) = \phi_0 + \delta\phi_0\, \xi(x,y)$, and $Q_{ij}(x,y,0) = \delta Q\, \xi(x,y)$, where $\xi(x,y)$ is a uniformly distributed random variable in the range $[-1,1]$, and $\delta\phi_0 = 0.01$ and $\delta Q = 0.01$ are small perturbation amplitudes. The phase field $\phi$ evolves according to the Cahn--Hilliard equation, discretized using second-order central finite differences, while the nematic order parameter $Q_{ij}$ is updated using the Beris--Edwards equation. To ensure numerical stability and accuracy, a sub-stepping procedure is employed in which five finite-difference updates are performed within each LBM time step, corresponding to an effective time step $\Delta t_{\mathrm{fd}} = \Delta t / 5$. The physical parameters are chosen within ranges commonly used in the literature: $\rho_1 = \rho_2 = 1.0$, $\eta_1 = \eta_2 = 0.5$, $M_\phi = 0.2$, $A = 0.05$, $\kappa = 0.1$, $\gamma = 0.1$, $\Gamma_1 = \Gamma_2 = 0.1$, $\Gamma_Q = 0.2$, and $K = 0.1$, while the activity parameter $\zeta$ is varied systematically. To validate the implementation, we consider the limiting case of a single-fluid active nematic system by setting $\phi = 1$, $\mathbf{u}_2 = 0$, and $\gamma = 0$, and we observe excellent agreement of the spatial velocity correlation functions with previously reported results, such as those of Thampi \textit{et al.} \cite{Thampi-PTSAMPES-2014} as shown in Fig.\ref{Fig01}, confirming the accuracy of the numerical approach.

\section{Results and Discussion}
We begin by analyzing the evolution of phase separation and flow structures from an initially homogeneous state with mean concentration $\phi_0 = 0.5$. The interplay between composition dynamics and active hydrodynamics leads to rich non-equilibrium behavior, which we characterize through both visual and statistical measures.

\begin{figure}[t]
\centering
\includegraphics[width=0.5\textwidth]{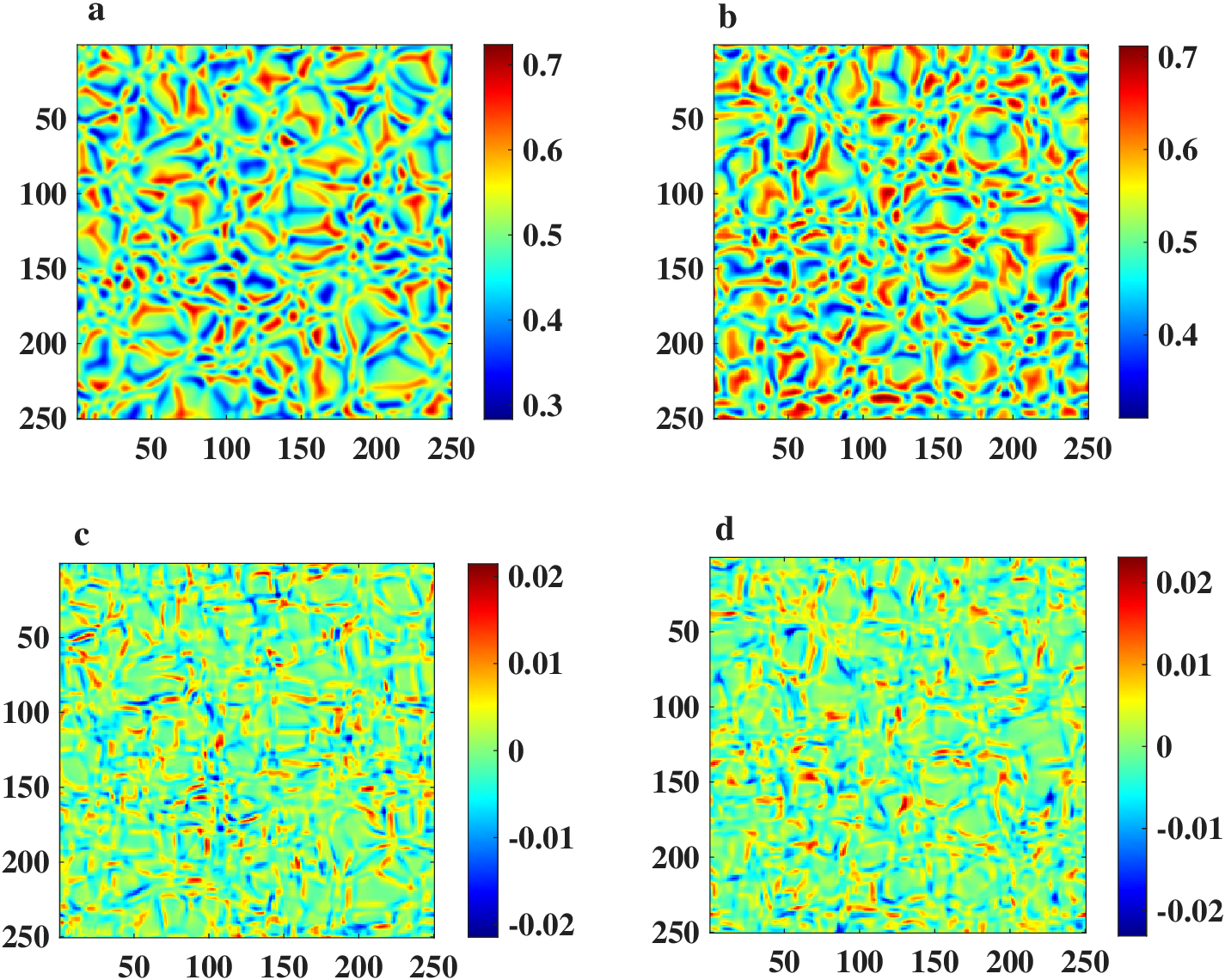}
\caption{Pseudocolor plots of the concentration field $\phi$ (top row) and the corresponding mixed vorticity field $\omega_c$ (bottom row) for different activity strengths $\zeta$. The mixed velocity is defined as ${\mathbf{u}}_c = \phi \mathbf{u}_1 + (1 - \phi)\mathbf{u}_2$, and the vorticity is computed as $\omega_c = \nabla \times {\mathbf{u}}_c$. Panels (a) and (b) show the concentration field for $\zeta = 0.001$ and $\zeta = 2.0$, respectively, while panels (c) and (d) display the corresponding vorticity fields.}
\label{Fig02}
\end{figure}

The mixed (barycentric) velocity of the system is defined as
\begin{equation}
{\mathbf{u}}_c = \phi \mathbf{u}_1 + (1 - \phi)\mathbf{u}_2,
\end{equation}
and the corresponding mixed vorticity field is
\begin{equation}
\omega_c = \nabla \times {\mathbf{u}}_c.
\end{equation}
In two dimensions, this reduces to
\begin{equation}
\omega_c = \frac{\partial u_{c,y}}{\partial x} - \frac{\partial u_{c,x}}{\partial y}.
\end{equation}
Fig.\ref{Fig02} shows representative snapshots of the concentration field and the corresponding vorticity field for different values of the activity parameter $\zeta$. At low activity ($\zeta = 0.001$), the system exhibits weak phase separation with small fluctuations and negligible flow structures. As the activity increases, the concentration domains become increasingly distorted and the vorticity field develops strong rotational structures, indicating the emergence of active turbulence. This behavior is consistent with recent studies of active-scalar systems, where activity injects energy into the flow and generates a statistically steady turbulent state. In such systems, the coupling between the scalar field and velocity field leads to a non-equilibrium steady state characterized by chaotic flow patterns and enhanced mixing. \cite{Doostmohammadi-NC-2018, Bhattacharyya-PRL-2023}
\begin{figure}[t]
\centering
\includegraphics[width=0.5\textwidth]{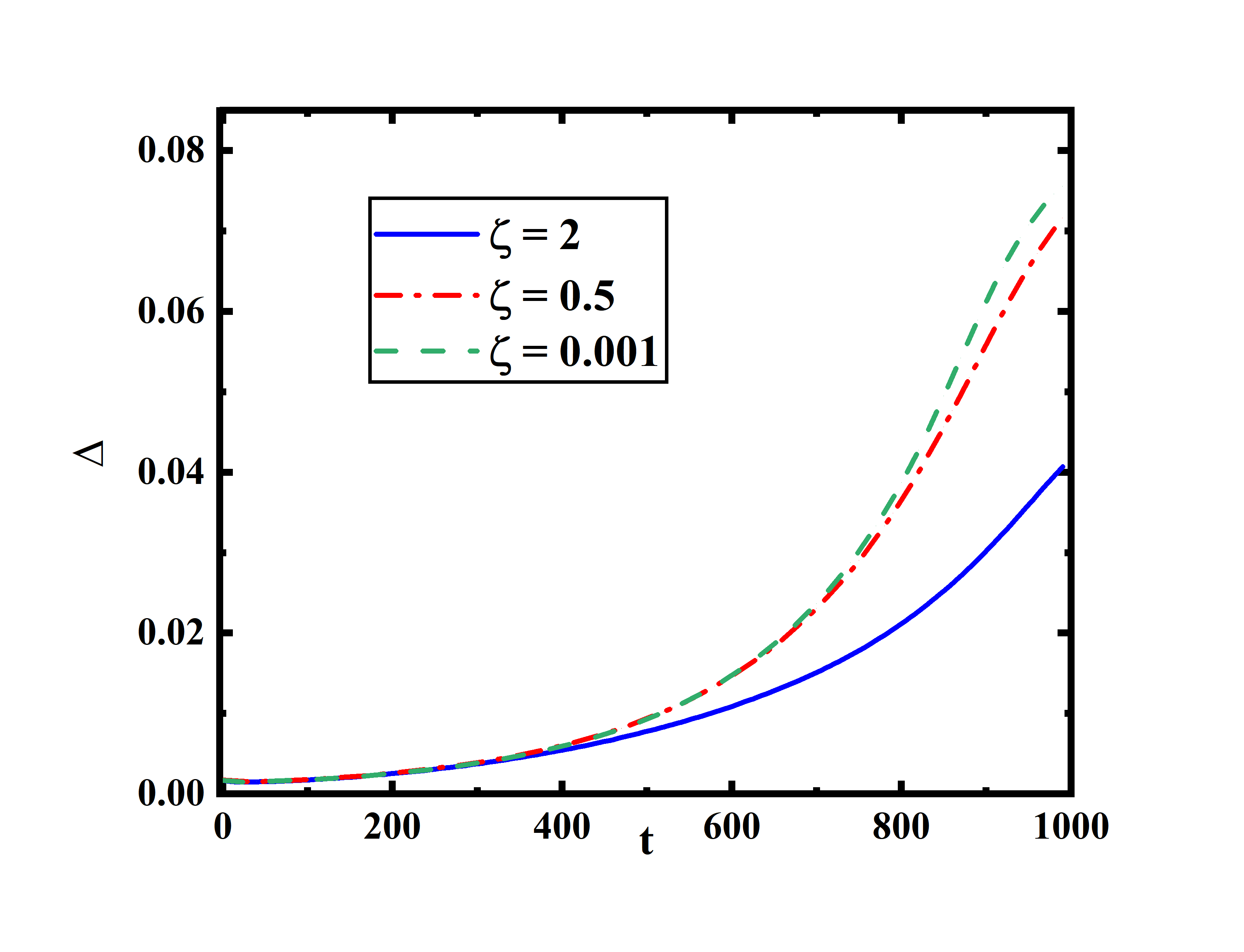}
\caption{Time evolution of the standard deviation of the concentration field, $\Delta$, for different values of the activity parameter $\zeta$.}
\label{Fig03}
\end{figure}
\begin{figure*}[t]
\centering
\subfloat[]{\includegraphics[width=0.5\textwidth]{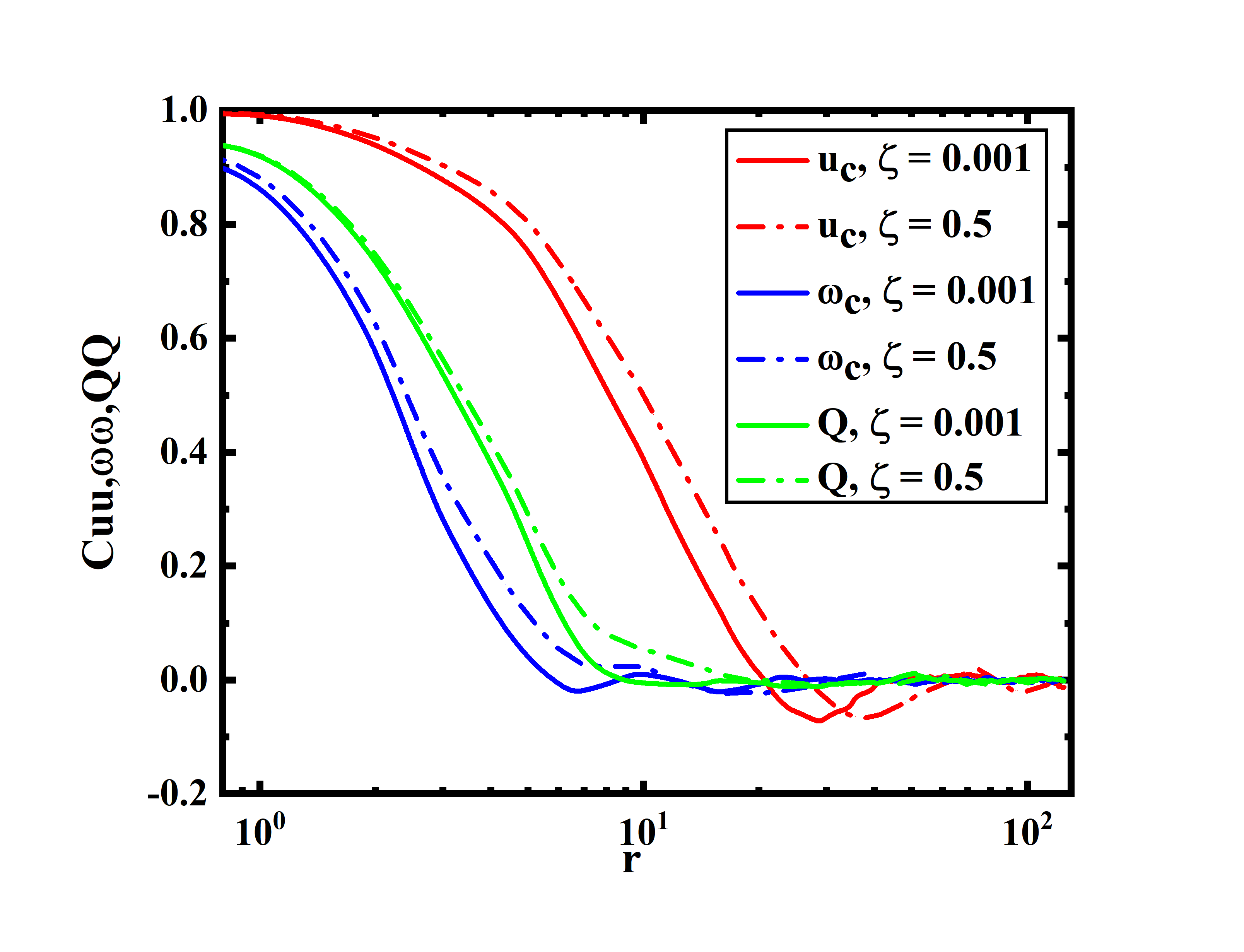}}
\subfloat[]{\includegraphics[width=0.5\textwidth]{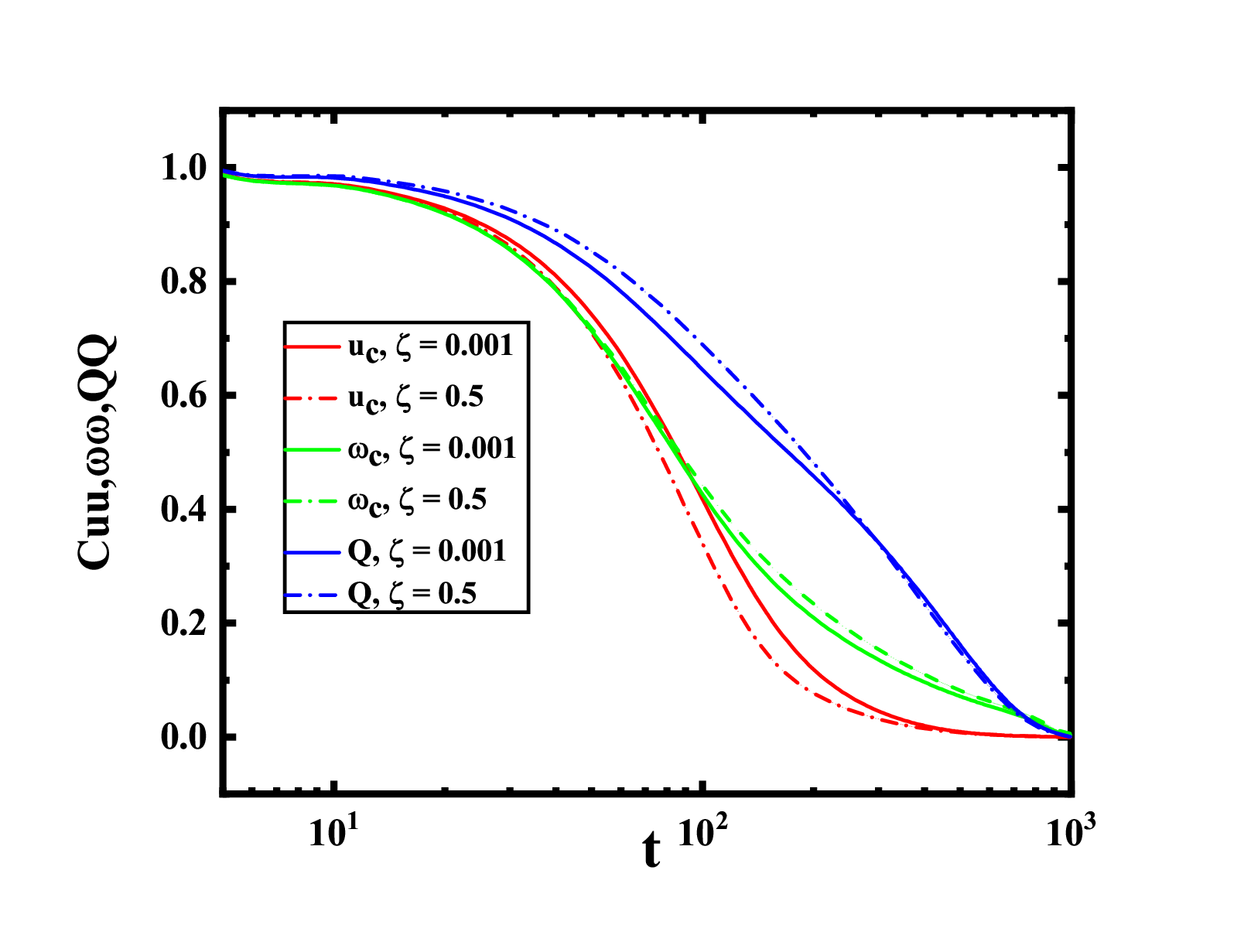}}
\caption{Spatial correlation functions $C{uu}$, $C\omega\omega$, and $CQQ$ versus separation distance $r$ for extensile activity strengths $\zeta = 0.001$ and $0.5$. (b) Temporal correlation functions  $C{uu}$, $C\omega\omega$, and $CQQ$ versus time lag $t$ for the same activity values. Colors distinguish the three quantities: blue for mixed velocity $u_c$, red for mixed vorticity $\omega_c$, and green for nematic tensor $\mathbf{Q}$. Within each color, different line styles (solid, dashed, dotted) correspond to $\zeta = 0.001$, and $0.5$, respectively. Increasing $\zeta$ leads to faster decay of both spatial and temporal correlations for all three quantities, indicating the transition from passive coarsening to active turbulence.}
\label{Fig04}
\end{figure*}
To quantify the degree of phase separation, we compute the standard deviation of the concentration field, $\Delta$, as shown in Fig.\ref{Fig03} Initially, $\Delta$ increases as domains form and grow due to phase separation. However, at long times, $\Delta$ saturates instead of increasing indefinitely. This saturation indicates that the coarsening process is arrested by activity-induced flows.

The observed coarsening arrest arises from the competition between phase separation, which promotes domain growth, and active turbulence, which enhances mixing and fragmentation. As a result, the system reaches a dynamically steady microphase-separated state with a finite characteristic length scale. Similar behavior has been reported in active Cahn--Hilliard--Navier--Stokes systems, where activity-driven turbulence suppresses domain coarsening and stabilizes finite-sized structures.\cite{Padhan-SM-2024}

Furthermore, increasing the activity parameter $\zeta$ enhances the strength of vorticity and accelerates the mixing process, leading to stronger suppression of large-scale domains. This demonstrates that activity not only drives flow but also plays a crucial role in controlling the structural organization of the system.

To quantitatively examine the spatial coherence and temporal persistence of the active turbulent flows, we compute the spatial correlation functions
\begin{align}
    Cuu(r) = \frac{\langle \delta \mathbf{u}_c(0) \cdot \delta \mathbf{u}_c(\mathbf{r}) \rangle}{\langle |\delta \mathbf{u}_c|^2 \rangle},\\
C\omega\omega(r) = \frac{\langle \delta \omega_c(0) \, \delta \omega_c(\mathbf{r}) \rangle}{\langle |\delta \omega_c|^2 \rangle}, \\
CQQ(r) = \frac{\langle \mathbf{Q}(0) : \mathbf{Q}(\mathbf{r}) \rangle}{\langle \|\mathbf{Q}\|^2 \rangle},
\end{align}
and the corresponding temporal correlation functions
\begin{align}
Cuu(t) = \frac{\langle \mathbf{u}_c(\tau) \cdot \mathbf{u}_c(\tau + t) \rangle}{\langle |\mathbf{u}_c(\tau)|^2 \rangle}, \\
C\omega\omega(t) = \frac{\langle \omega_c(\tau) \, \omega_c(\tau + t) \rangle}{\langle |\omega_c(\tau)|^2 \rangle}, \\
CQQ(t) = \frac{\langle \mathbf{Q}(\tau) : \mathbf{Q}(\tau + t) \rangle}{\langle \|\mathbf{Q}(\tau)\|^2 \rangle},
\end{align}
which quantify the persistence of flow structures and nematic order over time. 

Figure~\ref{Fig04} presents these correlation functions for three activity strengths: $\zeta = 0.001$ and $\zeta = 0.5$. In both panels, colors distinguish the three quantities: blue for the mixed velocity $u_c$, red for the mixed vorticity $\omega_c$, and green for the nematic tensor $\mathbf{Q}$. Within each color group, different line styles (solid, dashed, dotted) correspond to $\zeta = 0.001$, and $0.5$, respectively.

Panel (a) shows the spatial correlation functions $Cuu(r)$, $C{\omega\omega}(r)$, and $C{QQ}(r)$ versus separation distance $r$. For the weakest activity ($\zeta = 0.001$), all three correlations decay slowly, indicating large, coherent structures typical of passive phase separation. As $\zeta$ increases to $0.5$, the correlation functions decay more rapidly for all three quantities, reflecting the fragmentation of large domains and the emergence of smaller, turbulent structures. Among the three, $C{QQ}(r)$ (green curves) exhibits the longest correlation length at the activities considered, suggesting that nematic orientational order persists over larger distances than velocity or vorticity correlations, although the difference becomes less pronounced at $\zeta = 0.5$.

Panel (b) displays the temporal correlation functions $C{uu}(t)$, $C{\omega\omega}(t)$, and $C{QQ}(t)$. For $\zeta = 0.001$, all temporal correlations decay slowly, indicating persistent flow structures and orientational order with long memory. As activity increases to  $\zeta = 0.5$, the decay becomes significantly faster for all three quantities, demonstrating that stronger active turbulence leads to more rapid loss of temporal coherence. The decay rate of these temporal correlations provides a qualitative indication of the turbulent intensity: higher activity injects more energy into the system, producing more erratic and rapidly decorrelating flows. Notably, $C{QQ}(t)$ tends to decay more slowly than $C{uu}(t)$ and $C{\omega\omega}(t)$ for the activity values shown, indicating that the nematic field retains memory longer than the hydrodynamic fields under these conditions.

To further characterize the active turbulent state, we compute the isotropic kinetic energy spectrum of the mixed velocity field ${\mathbf{u}}_c$, the enstrophy spectrum, and the phase-field spectrum. Following the definitions in our numerical code, these are given by:
\begin{equation}
E_{kin}(k) = \frac{1}{2} \langle \hat{{\mathbf{u}}}_c(\mathbf{k}) \cdot \hat{{\mathbf{u}}}_c(-\mathbf{k}) \rangle,
\end{equation}
\begin{equation}
W(k) = k^2 E_{kin}(k),
\end{equation}
\begin{equation}
S^{\phi}(k) = \langle |\hat{\phi}(\mathbf{k})|^2 \rangle,
\end{equation}
where hats denote Fourier transforms, $k = |\mathbf{k}|$, and $\langle \cdot \rangle$ represents shell averaging over the statistically steady state.
\begin{figure*}[t]
\centering
\subfloat[]{\includegraphics[width=0.5\textwidth]{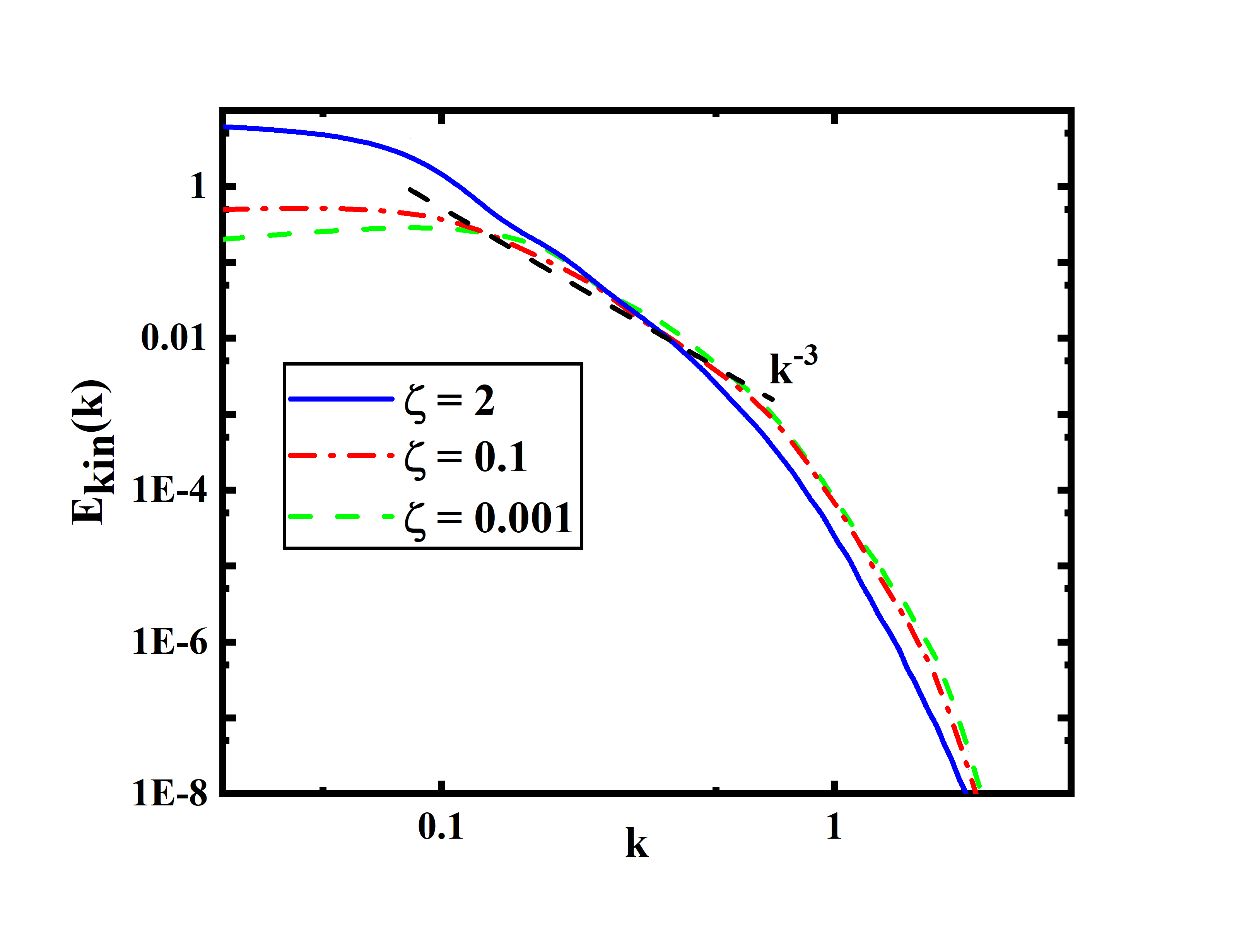}}
\subfloat[]{\includegraphics[width=0.5\textwidth]{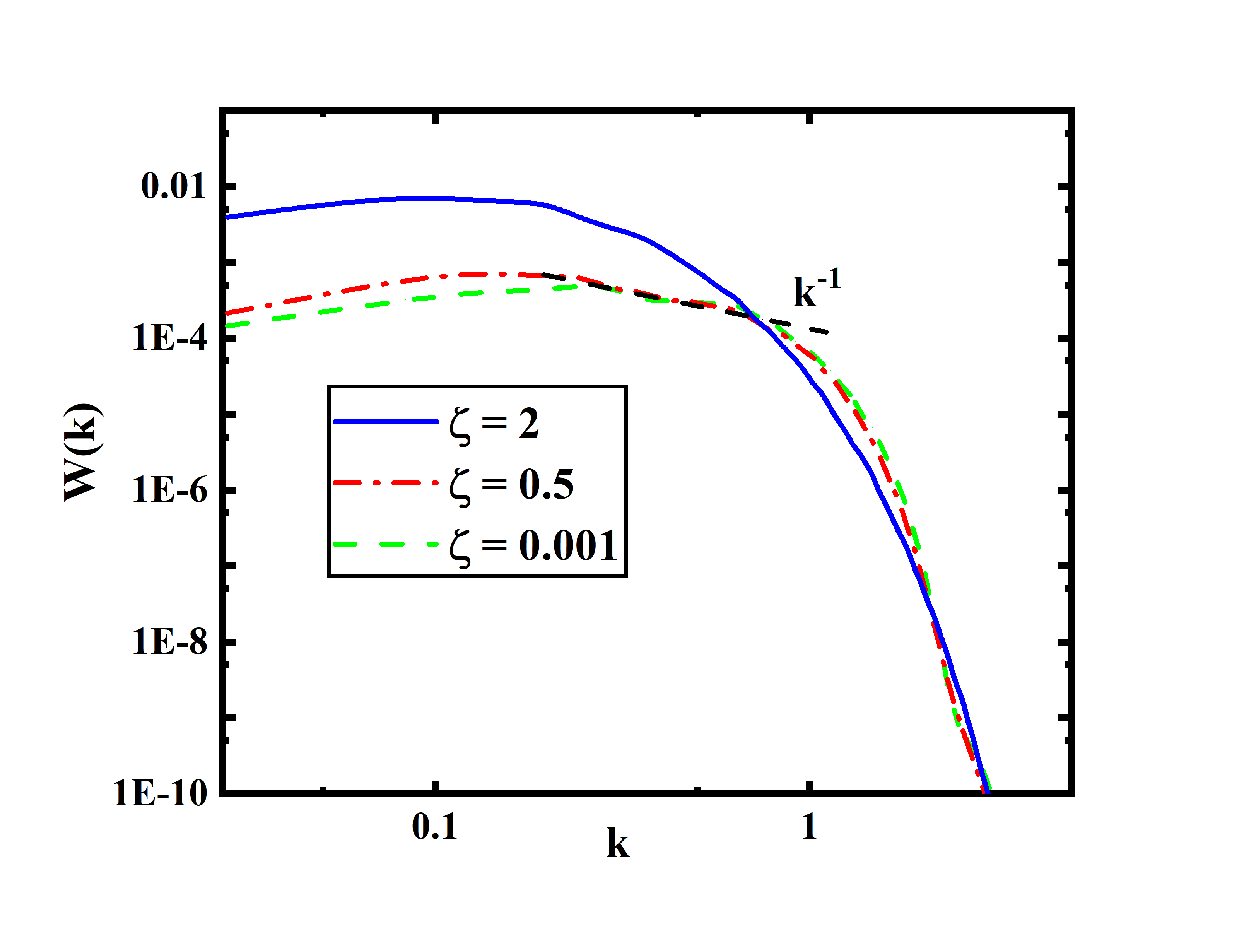}}\\
\subfloat[]{\includegraphics[width=0.5\textwidth]{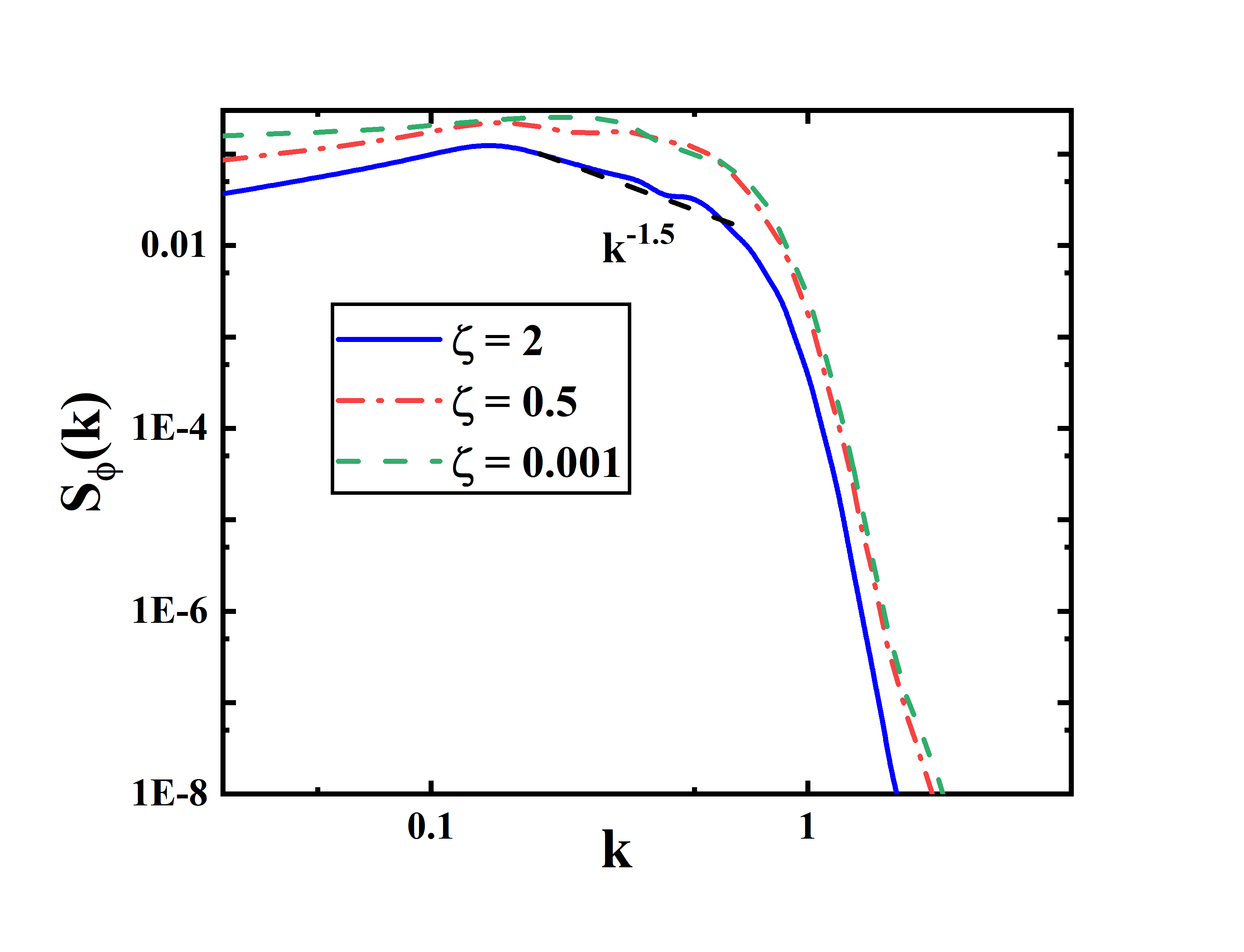}}
\caption{(a) Kinetic energy spectrum $E_{kin}(k)$ vs. $k$ for $\zeta = 0.001$, $0.5$, and $2.0$. (b) Enstrophy spectrum $W(k)=k^2E_{kin}(k)$. (c) Phase-field spectrum $S^{\phi}(k)$. For $\zeta=2.0$, the spectra follow $E_{kin}\sim k^{-3}$, $W\sim k^{-1}$, and $S^{\phi}\sim k^{-3/2}$, indicating two-dimensional active turbulence and coarsening arrest All simulations are in the low-Reynolds-number regime ($Re < 1$).}
\label{Fig05}
\end{figure*}
Fig.~\ref{Fig05} displays these spectra for three activity strengths: $\zeta = 0.001$ (weak activity), $\zeta = 0.5$ (moderate activity), and $\zeta = 2.0$ (strong activity). All spectra are plotted on logarithmic axes.

Panel (a) shows the kinetic energy spectrum $E_{kin}(k)$. At the lowest activity ($\zeta = 0.001$), the spectrum decays steeply at high wavenumbers, indicating that kinetic energy is concentrated at large scales, characteristic of passive coarsening without significant turbulent energy injection. As activity increases to $\zeta = 0.5$ and $\zeta = 2.0$, the spectra broaden and develop power‑law behavior. For $\zeta = 2.0$, $E_{kin}(k)$ exhibits a clear $k^{-3}$ scaling over approximately one decade of wavenumbers. $k^{-3}$ is commonly observed in active nematic turbulence and is associated with enstrophy-like transfer in small scale.

Panel (b) presents the enstrophy spectrum $W(k) = k^2 E_{kin}(k)$. For $\zeta = 2.0$, substituting $E_{kin}(k) \sim k^{-3}$ yields $W(k) \sim k^{-1}$, which is indeed observed. The observed $k^{-1}$ scaling follows directly from the relation $W(k)= k^{2}E(k^)$ and is consistent with the measured kinetic energy spectrum. For $\zeta = 0.5$, the enstrophy spectrum shows an intermediate behavior, while at $\zeta = 0.001$ it decays more steeply, indicating that enstrophy is concentrated at large scales without an inertial range.

Panel (c) displays the phase‑field spectrum $S^{\phi}(k)$, which quantifies concentration fluctuations. At $\zeta = 0.001$, $S^{\phi}(k)$ exhibits a pronounced peak at low wavenumbers, reflecting large, coarsening domains typical of passive phase separation. As activity increases to $\zeta = 0.5$ and $\zeta = 2.0$, the spectral weight shifts to higher wavenumbers, and the low‑wavenumber peak becomes less pronounced. For $\zeta = 2.0$, $S^{\phi}(k)$ shows a power‑law decay $S^{\phi}(k) \sim k^{-3/2}$ over an intermediate range of wavenumbers. Similar power‑law decays have been reported in active scalar turbulence \cite{Padhan-SM-2024}, where activity‑driven flows suppress large‑scale concentration fluctuations. This suggests a self‑similar distribution of concentration fluctuations under strong active mixing.

The shift of spectral weight to higher wavenumbers in $S^{\phi}(k)$ as $\zeta$ increases provides direct spectral evidence for coarsening arrest: active turbulence fragments large concentration domains, suppressing the growth of the characteristic length scale and stabilizing a microphase‑separated steady state.

It is important to note that all simulations operate at low Reynolds numbers ($Re < 1$), as calculated using $Re = \rho u_{rms} L_a / \eta$ with $L_a = \sqrt{K/\zeta}$. Despite the dominance of viscous effects, active stresses inject energy across scales, generating the observed power‑law spectra. This confirms that active turbulence is fundamentally different from classical inertial turbulence: it arises from internal active forcing rather than external inertial driving \cite{Doostmohammadi-NC-2018}.

Together, these spectral signatures provide quantitative evidence that increasing extensile activity drives the system from a passive coarsening regime into an active turbulent state characterized by a $k^{-3}$ kinetic energy spectrum (forward enstrophy cascade), a $k^{-1}$ enstrophy spectrum, and a $k^{-3/2}$ phase‑field spectrum (coarsening arrest). The emergence of these power laws at $Re < 1$ confirms that activity alone can sustain a turbulent cascade without the need for high inertia.

\begin{figure}[t]
\centering
\includegraphics[width=0.5\textwidth]{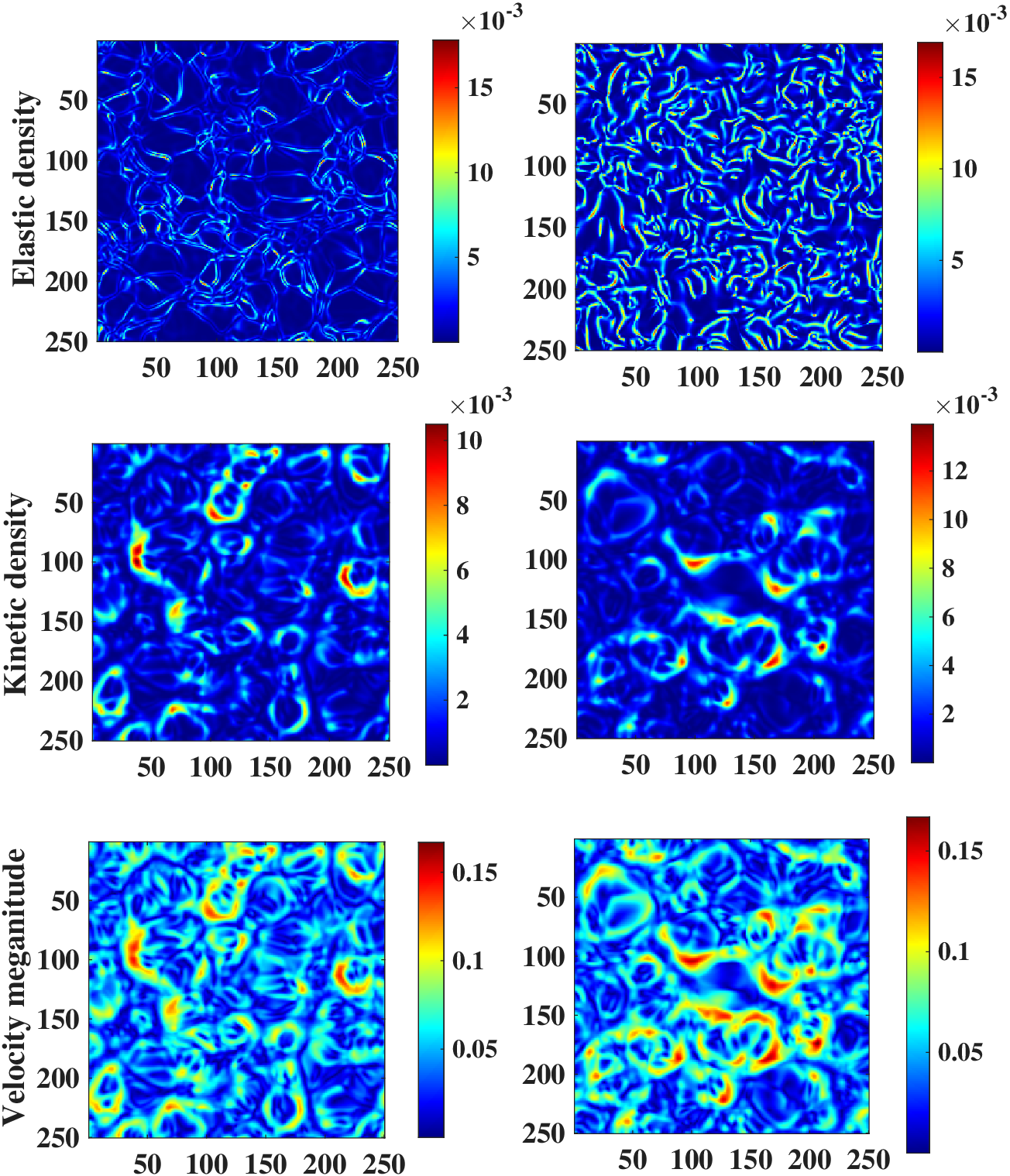}
\caption{Spatial distribution of nematic elastic energy density, kinetic energy density, and mixed velocity magnitude for different activity levels. The top row shows the elastic energy density, the middle row shows the kinetic energy density, and the bottom row shows the velocity magnitude for (left) $\zeta = 0.01$ and (right) $\zeta = 2.0$. At low activity ($\zeta = 0.01$), the system shows large, smooth structures with relatively uniform energy distributions. At higher activity ($\zeta = 2.0$), the system exhibits more fragmented, small-scale patterns, with stronger localized kinetic energy and highly concentrated elastic distortions, indicating active turbulence-like dynamics.}
\label{Fig06}
\end{figure}

To visualize how activity reorganizes the spatial distribution of energy and flow, Fig.\ref{Fig06} presents pseudocolor plots of three key quantities for weak ($\zeta = 0.01$) and strong ($\zeta = 2.0$) activity. The top row shows the nematic elastic energy density, defined as
\begin{equation}
E_{\mathrm{elastic}} = \frac{K}{2} |\nabla \mathbf{Q}|^2,
\end{equation}
which penalizes distortions in the nematic director field. The middle row displays the kinetic energy density of the mixed flow,
\begin{equation}
E_{\mathrm{kin}} = \frac{1}{2}\rho_c |{\mathbf{u}}_c|^2, 
\end{equation}
where $\rho_c = \rho_1 + \rho_2$ is the total density. The bottom row shows the magnitude of the mixed velocity field $|{\mathbf{u}}_c|$.

At low activity ($\zeta = 0.01$, left column), the elastic energy density is diffuse and concentrated in broad, interconnected regions, reflecting a weakly distorted nematic field. The kinetic energy density and velocity magnitude exhibit large-scale, coherent structures with smooth gradients, characteristic of passive phase separation where flow is driven primarily by interfacial tension rather than active stresses. The velocity field is weak and organized into large circulating patterns.

At high activity ($\zeta = 2.0$, right column), a dramatic transformation occurs. The elastic energy density becomes highly localized and fragmented, with intense spots corresponding to regions of strong nematic distortion. The regions of higher elastic energy density in the figure may correspond to regions of strong nematic distortions, which could be associated with defect cores or defect-rich regions. However, further quantitative analysis is required to directly confirm this association. The kinetic energy density and velocity magnitude reveal intense, small-scale vortical structures with sharp gradients, characteristic of active turbulence. The flow field is chaotic, with multiple interacting vortices spanning a range of scales.

The comparison between weak and strong activity demonstrates that active stresses inject energy into the system, fragmenting large-scale structures and generating turbulent fluctuations. The localization of elastic energy at high $\zeta$ indicates that the nematic field is strongly distorted by the chaotic flow, while the enhanced kinetic energy density confirms that activity drives vigorous fluid motion even at low Reynolds numbers ($Re < 1$). The spatial patterns in Fig.\ref{Fig06} provide visual evidence for the transition from passive coarsening to active-turbulence-arrested microphase separation, consistent with the spectral signatures shown in Fig.~\ref{Fig05}.

\begin{figure*}[t]
\centering
\subfloat[]{\includegraphics[width=0.5\textwidth]{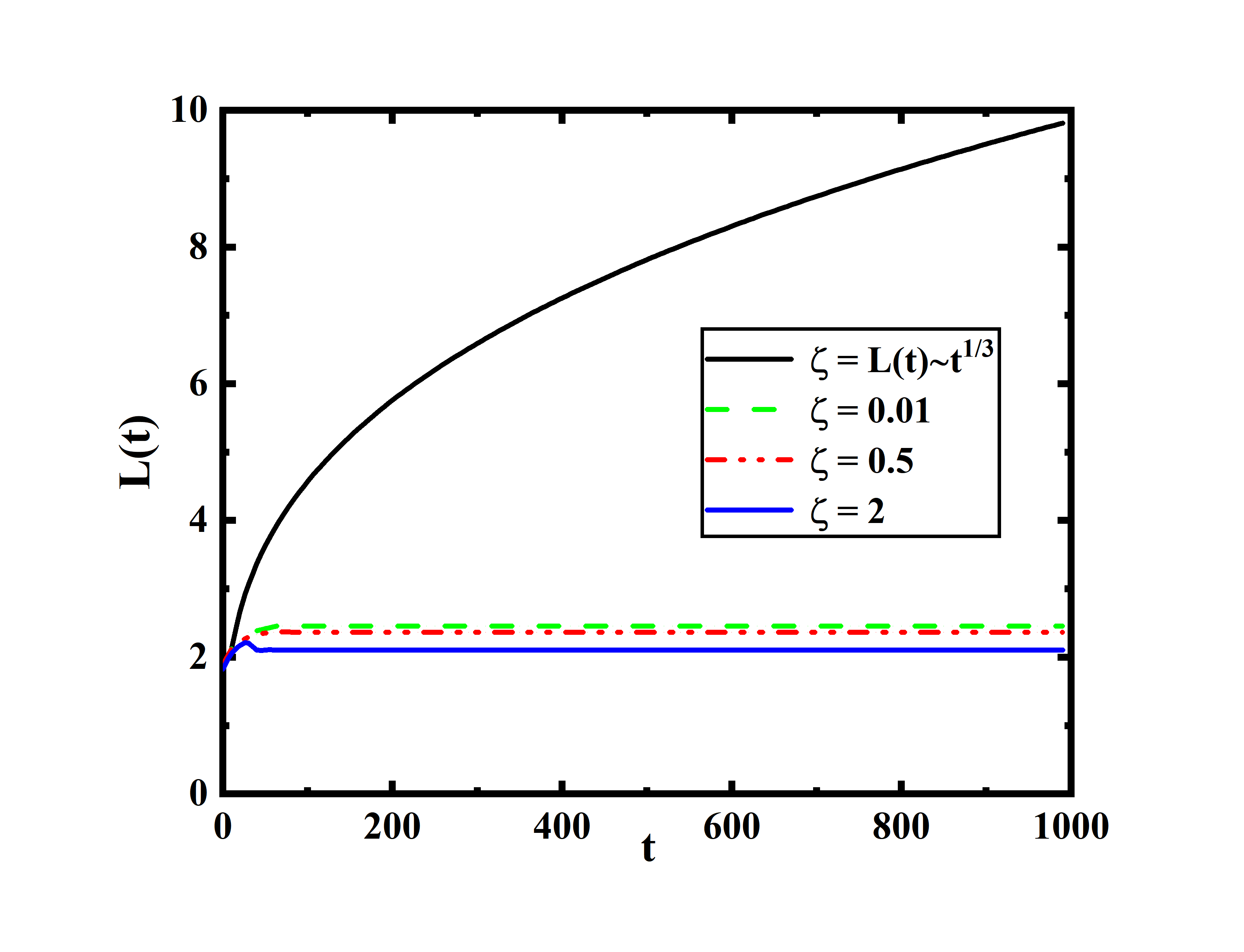}}
\subfloat[]{\includegraphics[width=0.5\textwidth]{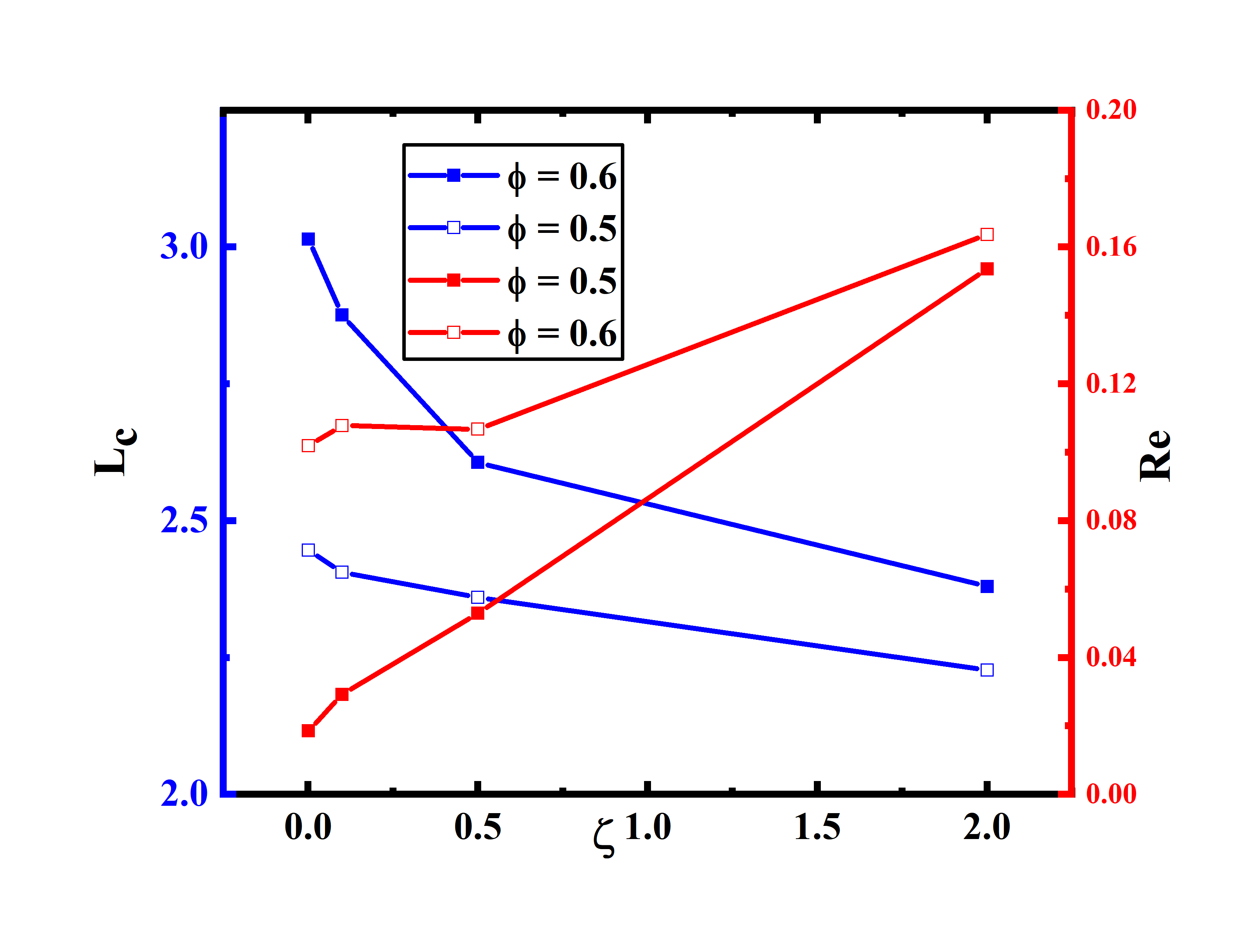}}\\
\subfloat[]{\includegraphics[width=0.5\textwidth]{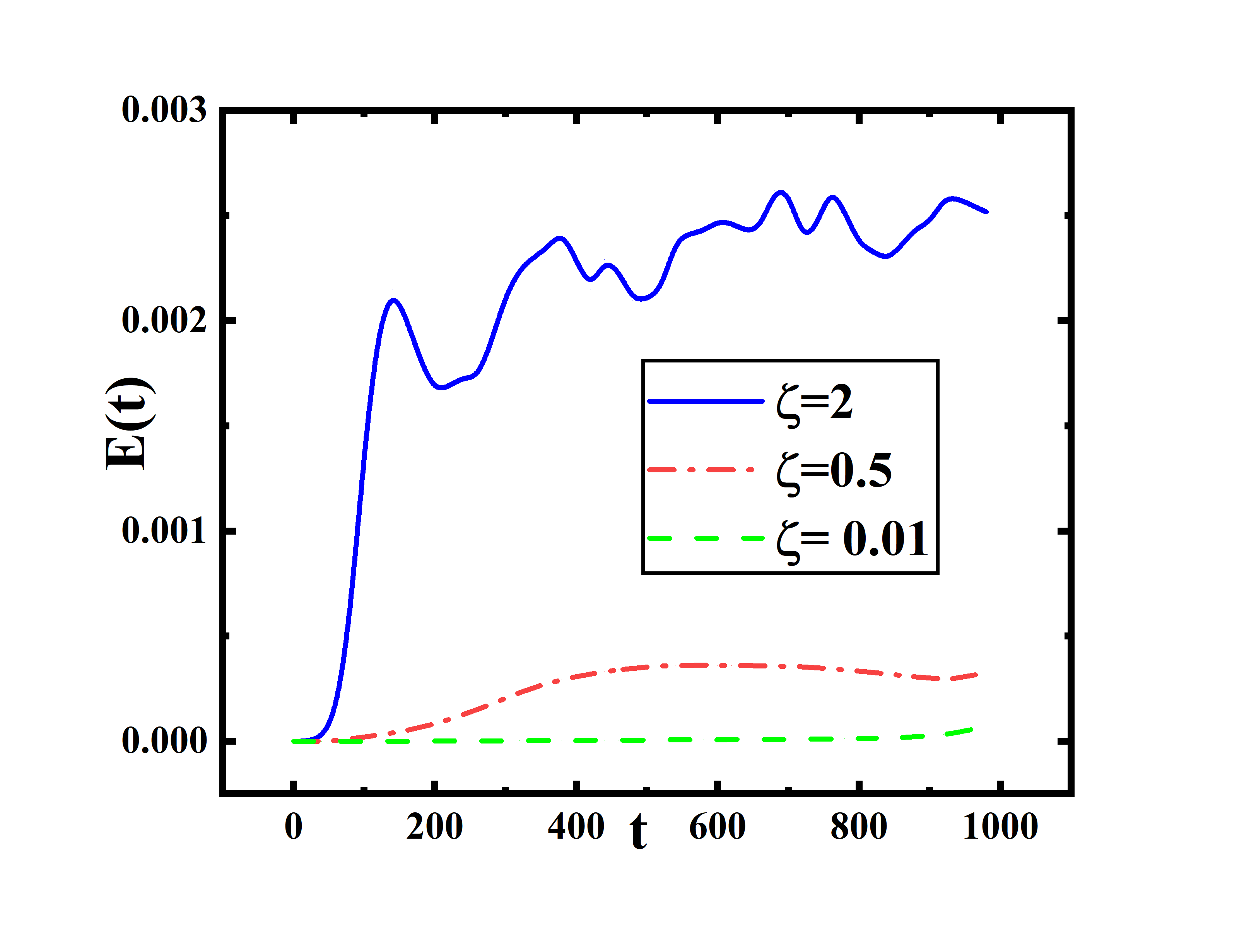}}
\caption{(a) Plot of the coarsening length scale $L(t)$ versus time $t$ for various values of $|\zeta|$; (b) plots of the mean coarsening-arrest scale Lc; (c) Plots of the kinetic energy density $E(t)$ versus time $t$ for various values of $\zeta$. }
\label{Fig07}
\end{figure*}

Fig.\ref{Fig07} presents the coarsening dynamics and the Reynolds number. Panel (a) tracks the temporal evolution of the coarsening length scale $L(t)$, defined from the structure factor of the concentration field as
\begin{equation}
L(t) = \frac{\sum_k \Phi(k,t)}{\sum_k k \Phi(k,t)}, \quad
\Phi(k,t) = \frac{1}{2} \sum_{k' = k-1/2}^{k' = k+1/2} |\hat{\phi}(k',t)|^2,
\end{equation}
for different activity strengths $\zeta$. For small times, $L(t)$ grows following a universal power law $L(t) \sim t^{1/3}$ (indicated by the solid guideline), independent of $\zeta$, consistent with diffusion‑limited (Lifshitz–Slyozov) coarsening. At later times, $L(t)$ saturates to a finite value for all $\zeta$, indicating coarsening arrest due to active turbulence.

\begin{figure*}[t]
\centering
\subfloat[]{\includegraphics[width=0.5\textwidth]{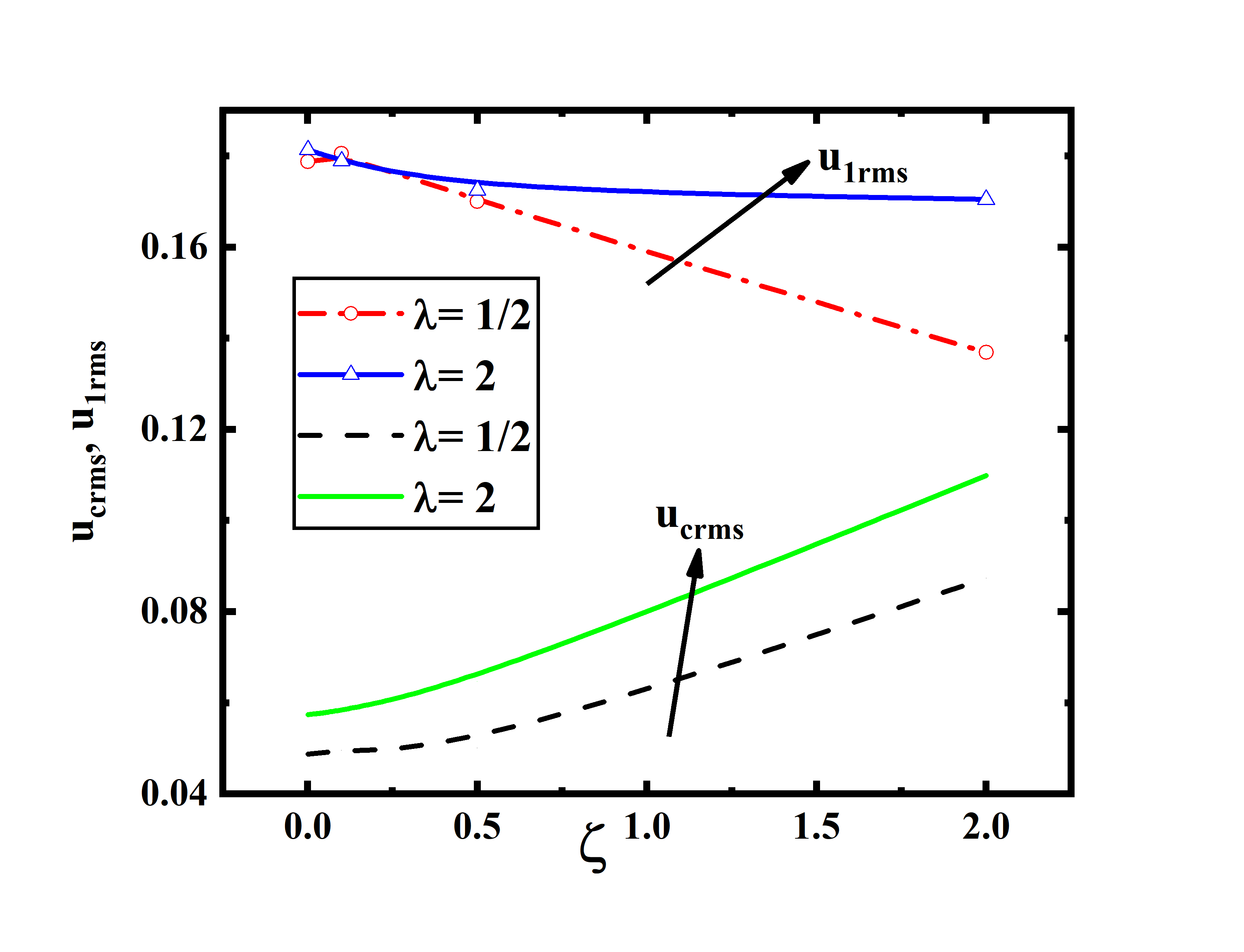}}
\subfloat[]{\includegraphics[width=0.5\textwidth]{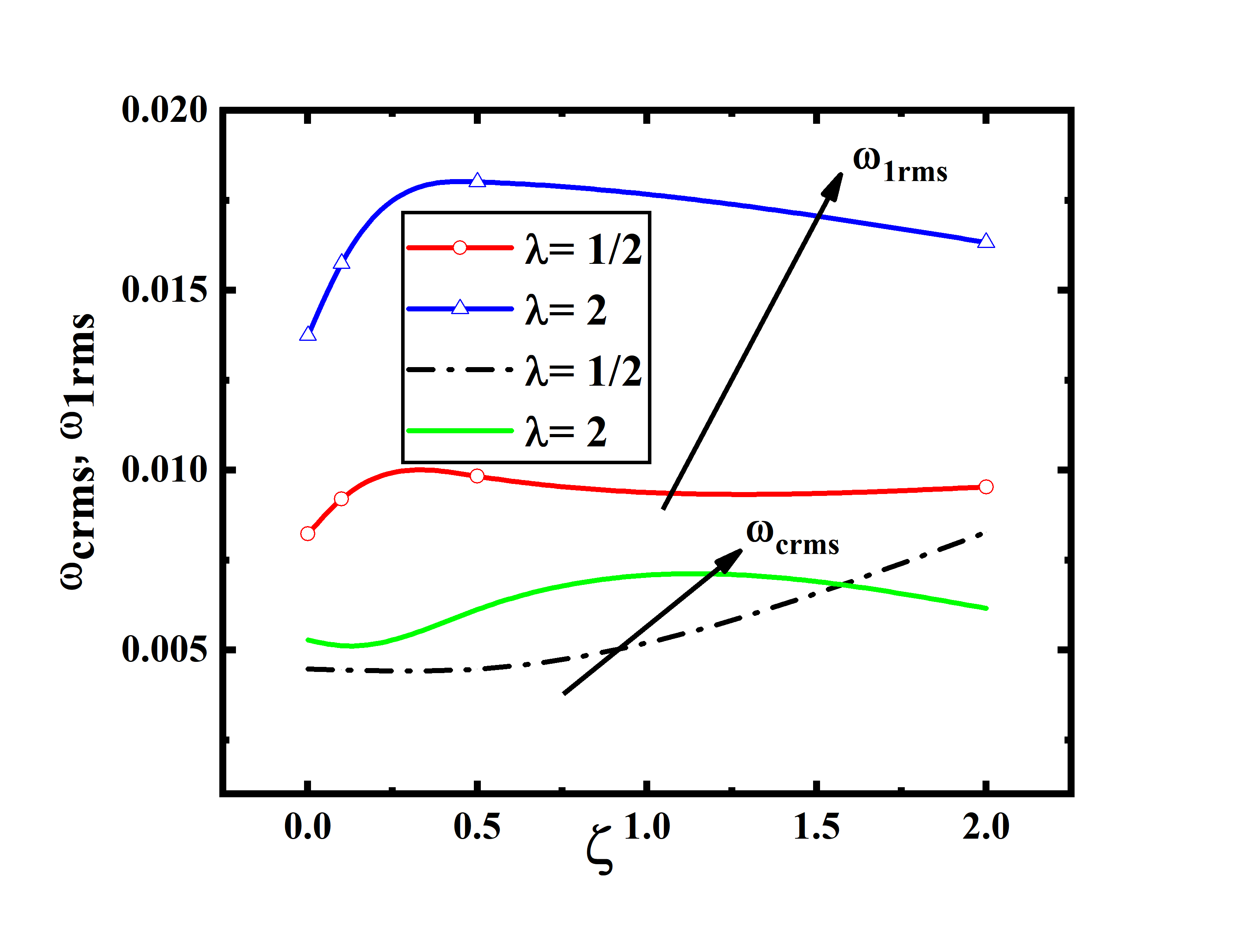}}
\caption{Root mean square velocity of active fluid $u_{1rms}$, mixture $u_{crms}$ (a) and vorticity of active fluid $\omega_{1rms}$, and mixture $\omega_{crms}$ (b)  for different values of active parameter $\zeta$  in both the flow-tumbling  ($\lambda=1/2$) and shear-aligning ($\lambda=2$) regimes.}\label{Fig08}
\end{figure*}
\begin{figure}[t]
\centering
\includegraphics[width=0.5\textwidth]{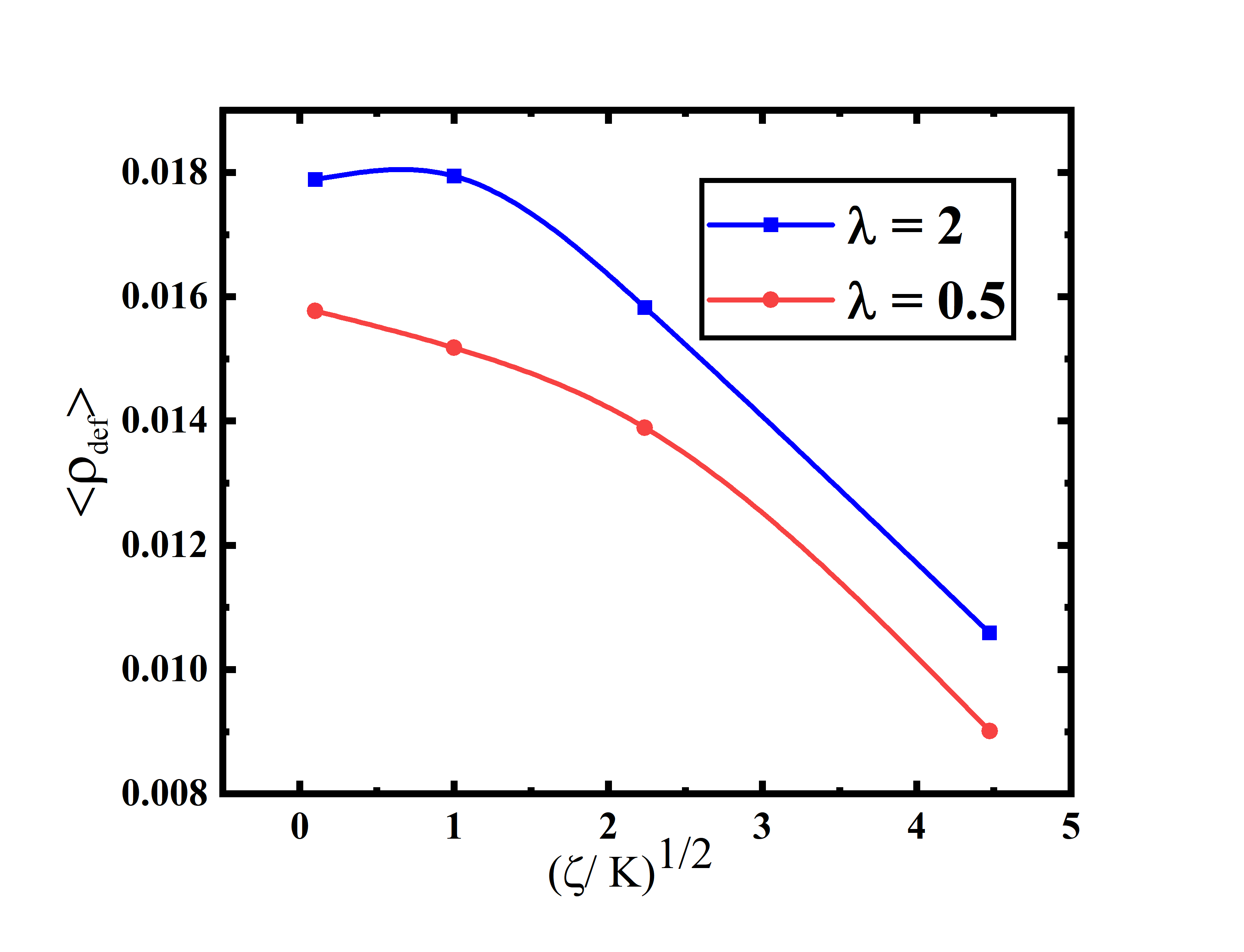}
\caption{Averaged defect density $\langle \rho_{\mathrm{def}}\rangle$ as a function of activity $\zeta$ for flow‑aligning ($\lambda = 2$) and flow‑tumbling ($\lambda = 0.5$) regimes at $\phi_0 = 0.5$. The shear‑aligning regime produces a consistently higher defect density, indicating stronger active turbulence.}\label{Fig09}
\end{figure}
Panel (b) shows the mean coarsening‑arrest scale $L_c = \langle L(t) \rangle_t$ (left axis) and the Reynolds number $Re$ (right axis) as functions of $\zeta$. The Reynolds number is defined as
\begin{equation}
Re = \frac{\rho_c \, u_{c,\mathrm{rms}} \, L_a}{\eta_c}, \qquad
L_a = \sqrt{\frac{K}{|\zeta|}},
\end{equation}
where $\rho_c = \rho_1 + \rho_2$ is the total density, $u_{c,\mathrm{rms}}$ is the root‑mean‑square velocity of the mixed flow, $L_a$ is the active length scale, and $\eta_c = \eta_1 + \eta_2$ is the total viscosity. Mean arrested length scale $L_c$ decreases and $Re$ increases with $\zeta$, indicating that stronger activity leads to smaller characteristic domain sizes in the arrested steady state and simultaneously drives more vigorous turbulent flow (higher $Re$).

Panel (c) plots the time evolution of the kinetic energy density of the mixture,
\begin{equation}
E(t) = \frac{1}{2} \sum_{k' = k-1/2}^{k' = k+1/2} |\hat{{\mathbf{u}}}_c(k',t)|^2,
\end{equation}
for various $\zeta$. At early times, $E(t)$ rises as active stresses inject energy. Both the growth rate and the saturated value increase with $\zeta$, confirming that higher activity leads to stronger steady‑state turbulence. The late‑time saturation reflects a balance between energy injection by active stresses and viscous dissipation is consistent with the statistically steady turbulent regime in active nematics. Although $Re < 1$, the system exhibits turbulence-like chaotic flows driven by internal active stresses rather than inertial nonlinearities.

The response of the nematic director to flow is governed by the tumbling parameter $\lambda$. For $|\lambda| < 1$ (here $\lambda = 1/2$) the director tumbles continuously under shear, whereas for $|\lambda| > 1$ (here $\lambda = 2$) it aligns at a steady angle relative to the flow direction. To investigate how this alignment affects the flow statistics, we compute the root‑mean‑square (RMS) velocity and vorticity of both the active fluid and the mixture.

Fig. \ref{Fig08} shows the RMS velocity of the active fluid $u_{1,\mathrm{rms}}$ (a) and of the mixture $u_{c,\mathrm{rms}}$ (b), together with the corresponding RMS vorticities $\omega_{1,\mathrm{rms}}$ (c) and $\omega_{c,\mathrm{rms}}$ (d), as functions of the activity strength $\zeta$ for the flow‑tumbling regime ($\lambda = 1/2$, dashed lines) and the shear‑aligning regime ($\lambda = 2$, solid lines). All quantities are taken from the statistically steady state.

For all four quantities, increasing $\zeta$ leads to larger RMS values, reflecting that stronger active stresses inject more kinetic energy and generate more intense vortical motion. The shear‑aligning regime ($\lambda = 2$) consistently produces higher RMS velocities and vorticities than the tumbling regime ($\lambda = 1/2$) at the same $\zeta$. This indicates that when the director aligns with the flow, it couples more efficiently to the velocity gradient, thereby enhancing the active forcing. In the tumbling regime, continuous director rotation reduces net alignment and weakens the transfer of energy from active stresses to fluid motion.

The difference between the two regimes is more pronounced for the active fluid quantities ($u_{1,\mathrm{rms}}$ and $\omega_{1,\mathrm{rms}}$) than for the mixture quantities ($u_{c,\mathrm{rms}}$ and $\omega_{c,\mathrm{rms}}$). This is expected because the active stress acts directly on the active fluid; the mixture velocity averages over both components, which partially smooths out the fluctuations. Nevertheless, the mixture vorticities also increase clearly with $\zeta$ and show a distinct dependence on $\lambda$, confirming that the nematic alignment regime influences the overall turbulent state of the two‑fluid system.

Topological defects play a central role in active nematics, as they are sites of intense flow and stress. We identify $\pm 1/2$ defects in the nematic director field using the winding number method and compute the averaged defect density $\langle \rho_{\mathrm{def}} \rangle$ as a function of the activity strength $\zeta$ for the two nematic alignment regimes: flow‑tumbling ($\lambda = 1/2$) and shear‑aligning ($\lambda = 2$). The results are shown in Fig.~\ref{Fig09} for a fixed mean concentration $\phi_0 = 0.5$.

For both regimes, $\langle \rho_{\mathrm{def}} \rangle$ increases with $\zeta$, indicating that stronger active stresses generate more topological defects. This is consistent with the picture of active turbulence, where defect proliferation is associated with the chaotic, highly vortical state. The shear‑aligning regime ($\lambda = 2$, solid line) exhibits a higher defect density than the tumbling regime ($\lambda = 1/2$, dashed line) across the entire range of $\zeta$. This difference arises because in the shear‑aligning case the director couples more strongly to the flow, leading to more efficient energy injection and hence a more turbulent state with a larger number of defects.

At the lowest activity ($\zeta = 0.01$), the defect density is very small in both regimes, as the system is nearly quiescent. As $\zeta$ increases, $\langle \rho_{\mathrm{def}} \rangle$ rises approximately monotonically, with the shear‑aligning curve remaining consistently above the tumbling curve. The gap between the two regimes widens with increasing $\zeta$, reflecting the growing influence of the alignment parameter on the nonlinear dynamics of active turbulence.
\section{CONCLUSIONS}
In this work, we developed a two-fluid continuum model for active–passive binary mixtures, coupling Cahn–Hilliard phase separation with full Beris–Edwards nematohydrodynamics and distinct momentum equations linked through viscous drag. This framework resolves relative motion and momentum transfer between active and passive components, extending beyond conventional single-fluid descriptions.

Our simulations demonstrate that activity fundamentally alters phase separation dynamics. While diffusion-driven coarsening initially follows the classical $L(t)\sim t^{1/3}$ scaling, active stresses introduce strong mixing and fragmentation, leading to coarsening arrest and the emergence of a steady microphase-separated state. The characteristic domain size decreases with increasing $\zeta$, highlighting the dominant role of active forcing over interfacial relaxation.

Spectral analysis reveals that the active turbulent state exhibits non‑trivial scaling behavior. In the low‑wavenumber range, the kinetic energy spectrum follows $E_{\rm kin}(k)\sim k^{-3}$, consistent with the forward enstrophy cascade of two‑dimensional turbulence. Consequently, the enstrophy spectrum scales as $W(k)=k^2E_{\rm kin}(k)\sim k^{-1}$. The phase‑field spectrum displays an approximate $S_{\phi}(k)\sim k^{-3/2}$ decay, indicating scale‑invariant concentration fluctuations under active mixing. These results demonstrate that activity can generate turbulence‑like cascades even in the low‑Reynolds‑number regime.

Correlation analysis further shows that increasing activity reduces both spatial and temporal coherence, leading to rapidly decorrelating velocity and vorticity fields. The nematic field retains comparatively longer correlations, reflecting the influence of elastic interactions. Moreover, the tumbling parameter $\lambda$ plays a key role: shear‑aligning systems ($|\lambda|>1$) exhibit stronger flows, higher root‑mean‑square velocities and vorticities, and enhanced defect proliferation compared to tumbling systems ($|\lambda|<1$).

Topological defects emerge as central dynamical structures in the active regime. Their density increases monotonically with activity, confirming that defect proliferation is closely linked to the transition toward active turbulence. The coupling between defect dynamics, flow generation, and phase separation underscores the complex interplay between hydrodynamics and nematic order in multi‑phase active systems.

Overall, this work highlights how activity, interphase drag, and nematic alignment collectively control flow structures, spectral energy distribution, and domain morphology in active binary mixtures. The proposed two‑fluid framework provides a versatile platform for studying realistic systems such as bacterial suspensions in complex fluids, active emulsions, and biological mixtures where relative motion between components is essential.
\section*{Acknowledgements}
We acknowledge the National Natural Science Foundation of China (NSFC) for the support our work through Grant No. 12202275.

\bibliography{apssamp}

\end{document}